# Origin of New Broad Raman D and G Peaks in Annealed Graphene

Jinpyo Hong,[†] Min Kyu Park,[†] Eun Jung Lee,[†] DaeEung Lee, Dong Seok Hwang and Sunmin Ryu[*]

Department of Applied Chemistry, Kyung Hee University, 1 Seocheon, Yongin, Gyeonggi 446-701, Korea

*E-mail: sunryu@khu.ac.kr

[†]These authors contributed equally to this work

**ABSTRACT:** Since graphene, a single sheet of graphite, has all of its carbon atoms on the surface, its property is very sensitive to materials contacting the surface. Herein, we report novel Raman peaks observed in annealed graphene and elucidate their chemical origins by Raman spectroscopy and atomic force microscopy (AFM). Graphene annealed in oxygen-free atmosphere revealed very broad additional Raman peaks overlapping the D, G and 2D peaks of graphene itself. Based on the topographic confirmation by AFM, the new Raman peaks were attributed to amorphous carbon formed on the surface of graphene by carbonization of environmental hydrocarbons. While the carbonaceous layers were formed for a wide range of annealing temperature and time, they could be effectively removed by prolonged annealing in vacuum. This study underlines that spectral features of graphene and presumably other 2-dimensional materials are highly vulnerable to interference by foreign materials of molecular thickness.

Since its first mechanical isolation from graphite in 2004,[1] graphene has been a subject of more than 20,000 research papers revealing a variety of its extraordinary materials properties. Such a drastic expansion has been boosted not only by various simple methods for exfoliation[1,2] and synthesis,[3] but also by the stability of the resulting graphene samples under various external perturbations.[4] Like graphite, graphene is chemically inert that treatments with most chemicals cause no structural damage and functionalization at the basal plane requires either mineral acids in the presence of strong oxidants[2] or radical species.[5,6] Graphene also remains intact at elevated temperatures in oxygen-free atmosphere.[7] The thermal stability, in particular, has allowed heating graphene up to several hundred degree C without severe degradation in preparing various types of graphene devices,[1] cleaning surfaces for scanning probe microscopy[8,9] and molecular assembly,[10] and chemical vapor deposition (CVD) of hexagonal BN with graphene as substrates.[11]

Despite the apparent thermal stability, however, thermal treatments cause graphene non-negligible changes in its material properties. Graphene supported on $SiO_2$/Si substrates, for example, undergoes in-plane compression[12,13] and rippling[14] due to conformal adhesion[15] or differential thermal expansion[13,16] of graphene with respect to the substrates upon annealing at >100 ºC. At the same time with the compression, annealed graphene becomes strongly p-doped by the ambient oxygen molecules.[14] As a result of both structural and electronic modifications, the charge carrier mobility of annealed graphene devices is much lower than that of pristine devices.[17] For graphene samples prepared by the mechanical exfoliation method using adhesive tape, heating is likely to induce contamination of graphene with nearby polymeric adhesive residues which may become fluidic or airborne when thermally decomposed.[7] Similar phenomenon may occur in CVD-grown graphene transferred onto solid substrates using polymeric supports which cannot be removed completely by solvent cleaning.[10] Notwithstanding its practical importance in understanding and improving properties of graphene sheets and devices, the annealing-induced contamination of graphene surface has not been investigated in a systematic manner.

Herein we report that novel broad Raman peaks emerge when graphene is annealed in oxygen-free environment and reveal that they originate not from graphene itself but from amorphous carbon (aC) generated from residual hydrocarbons at elevated temperature. This study also provides a straightforward optical method to monitor the surface quality of graphene, bearing practical importance towards many related fields.

**Results**

Figure 1a and 1b show the optical micrographs of exfoliated single layer graphene (denoted by 1L, spanning ~17x14 μm$^2$), obtained before and after thermal annealing in Ar atmosphere ($T_{anneal}$ = 400 ºC; $\tau_{anneal}$ = 2 hrs). (See experimental section for detailed methods.) Whereas graphene and thick graphite in yellow color remain optically identical after the annealing, adhesive residue marked by the yellow arrow disappeared through thermal decomposition of the polymeric materials. The Raman spectrum obtained from graphene in its pristine state (red in Fig.. 1c) shows only one first-order Raman peak at ~1580 cm$^{-1}$, denoted by G, which originates from the doubly degenerate zone-center phonon $E_{2g}$ mode and corresponds to the C-C stretching mode.[18] The D peak (~1340 cm$^{-1}$) arising from the TO phonon mode near K points in the Brillouin zone is hardly detectible in the pristine graphene. Since it is only activated by structural defects by a second order Raman scattering process through the intervalley double resonance, lack of the peak represents high crystallinity of the sample.[18] Graphene with significant defects would show an additional disorder related peaks, D' and (D+D') at ~1620 and ~2950 cm$^{-1}$, originating from the intravalley double resonance and a combination mode, respectively.[5,6] The second order Raman scattering in the pristine graphene also generated 2D and 2D' peaks at 2680 and 3240 cm$^{-1}$, respectively, overtones of D and D' modes.[19] Most of the pristine samples including the one in Fig. 1 showed D/G peak height ratio ($I_D/I_G$) of ~1% confirming high crystallinity,[7] while few with unusually high values were discarded to avoid complication.

When annealed for 2 hours at 300 °C in Ar gas, the spectrum shows decrease in 2D's intensity and upshifts in the frequencies of G and 2D, both of which are mostly due to annealing-induced hole doping caused by ambient $O_2$ molecules[13,14] with minor contribution from lattice compression.[13] We also note that the spectrum background between 1100 and 1600 cm$^{-1}$ increased noticeably. The spectral changes are more obvious in the expanded viewgraph in Fig. 1d. For the sample annealed at 400 °C, for example, the broad peak is clearly seen to be centered near the D peak. A close look further reveals that broad features also exist around G and 2D peaks, respectively. In particular, the spectrum near 1600 cm$^{-1}$ can be decomposed into a sharp G and a broad peak, located at 1603 and 1587 cm$^{-1}$, respectively (see Supplementary Fig. S1 online). The broad Raman peaks near D, G and 2D regions are strongly reminiscent of $sp^2$-rich amorphous carbons.[20,21] Indeed, the broad spectral features match well with the spectrum obtained from commercial carbon black powder, a form of amorphous carbons (see Supplementary Fig. S2). Thus we attribute the annealing-induced broad Raman peaks, denoted by $D_{aC}$, $G_{aC}$, $2D_{aC}$ and $(D+D')_{aC}$ to amorphous carbons (aC) deposited on graphene through carbonization[22] of hydrocarbons during the thermal treatments in inert atmosphere. Since the broad Raman peaks were not observed on the bare substrates near graphene flakes as explained below, it can be concluded that the adsorption of hydrocarbons and their carbonization occur selectively on the surface of graphene. The hydrocarbons, carbon source of aC, are thought to be the adhesive polymers left on the substrates during the mechanical exfoliation of graphene, since the polymeric residues were apparently removed by annealing and no other carbon source was available. We also found that the aC-derived Raman peaks occur in CVD-grown graphene transferred onto $SiO_2$/Si substrates using poly(methyl methacrylate) (PMMA) supports.[23] Since removal of PMMA by acetone is known to be incomplete,[10,24] these peaks are attributed to PMMA residues on graphene. (See Supplementary Figs. S3 and S4 online for detailed spectra and ensuing discussion.)

In Fig. 2a, we investigated the effects of the gas environment in the furnace quartz tube on the intensity of $D_{aC}$, $I(D_{aC})$, of exfoliated graphene. Since $D_{aC}$ cannot be separated unambiguously from the rest Raman features, $I(D_{aC})$ was arbitrarily defined as the intensity at 1280 cm$^{-1}$ to avoid overlap with the D peak from graphene itself. For quantitative comparison, all the spectra were normalized by the intensity of the Raman peak located at ~960 cm$^{-1}$ which originates from underlying Si substrates. Whereas $I(D_{aC})$ reached ~15% of the Si's reference peak when annealed in inert Ar atmosphere or reducing forming gas (5% $H_2$/95% $N_2$) at 400 °C, it decreased down to ~8% in a low vacuum (~2x10$^{-3}$ Torr). As decreasing the base pressure to a high vacuum level, $I(D_{aC})$ was further attenuated. This suggests that gaseous hydrocarbons emitted from the polymers at elevated temperature are lingering in the reaction chamber and then adsorb on graphene followed by carbonization to form aC.

The duration of annealing ($\tau_{anneal}$) was also found to affect $I(D_{aC})$ as shown in Fig. 2b. $I(D_{aC})$ peaked at the smallest $\tau_{anneal}$ of 0.5 hour in a low vacuum and decreased to a half for $\tau_{anneal}$ = 2 hours. Extended thermal treatment led to almost zero intensity of $D_{aC}$, suggesting that the surface of graphene is almost free of aC. Note that nonzero values for pristine graphene, $I(D_{aC})$ ~ 1.5%, is mostly due to unidentified optical scattering from the underlying substrates, not graphene or aC as can be seen in Fig. 1d. When $T_{anneal}$ was varied for exfoliated graphene, $I(D_{aC})$ started to increase at 300 °C and reached 7.2% ± 2.4% at 400 °C as shown in Fig. 2c. At even higher temperature (>800 °C), however, $I(D_{aC})$ decreased within increasing $T_{anneal}$. Rather large error bars in $I(D_{aC})$ are due to sample-to-sample variation, which possibly originates from random distribution of adhesive residues around graphene flakes of interest. A similar $T_{anneal}$-dependence was observed for CVD-grown graphene but with $I(D_{aC})$ overall higher than that for exfoliated graphene.

Using AFM, we were able to visualize overlayers of aC on graphene formed in Ar atmosphere at 700 °C. For this characterization, bilayer graphene (2L) shown in Fig. 3a was employed since 1L graphene is severely etched by a carbothermal reduction of underlying $SiO_2$.[25] To expose $SiO_2$ surface within the graphene basal plane as a height reference, etch pits were generated by oxidizing[7] graphene at 550 °C as shown in Fig. 3b. Then, the sample was annealed in Ar gas at 700 °C repeatedly for $\tau_{anneal}$ = 1, 2, 4 and 8 hours with the AFM images in Figs. 3b~3f obtained after each thermal treatment. Most of large pits (marked by the yellow arrow, diameter > 100 nm) shown in Fig. 3b turned out to grow into polygonal shapes for $\tau_{anneal}$ = 8 hours (Fig. 3f) through the carbothermal reduction,[25] indicating that they are 2L-deep pits[7] exposing the $SiO_2$ substrate. On the contrary, smaller pits of 1L depth[7] formed on the top graphene layer (encircled by the dotted line in Fig. 3b) remained same in size after 8-hour annealing (Fig. 3f) since the carbothermal reduction requires contact between $SiO_2$ and graphene edges.[25] Besides the anisotropic etching, the AFM images also reveal the presence and transformation of thin layers on top of graphene. Whereas the surface of oxidized graphene is flat except the pits (Fig. 3b), the subsequent one-hour annealing led to rather rugged surface (Fig. 3c) and additional treatment generated nicely stepped surface in Fig. 3d. The height profile averaged the yellow rectangle in Fig. 3d shows that the thickness of the overlayers is 0.5 ~ 0.7 nm (Fig. 3g). The third round of annealing removed most of the overlayers except those covering the small 1L-deep pits as shown in Fig. 3e. Finally, annealing for accumulated duration of 8 hours completely cleaned the surface as can be seen Fig. 3f. (See Supplementary Fig. S5 online for detailed height profile change for each of the AFM image.) Figure 3h presents $I(D_{aC})$ obtained after each thermal treatment. The $D_{aC}$ peak was most intense for $\tau_{anneal}$ = 1 hour and almost disappeared for $\tau_{anneal} \geq 4$ hours, showing a good correlation with the topographical details of the overlayers revealed by the AFM images. Comparison of Fig. 2b and Fig. 3h leads us to a conclusion that prolonged annealing at 700 °C is more efficient in removing the aC layers than at 400 °C.

Interestingly, the morphology of the aC overlayers was found to be very different when annealed at 400 °C in low vacuum. Instead of the planar structures observed in Fig. 3, the graphene annealed for 30 min is now covered with numerous tiny dots, which disappeared almost completely when annealed for 8 hours (see Supplementary Fig. S6 online). The height distribution of the 8-hour annealed graphene shown in Supplementary Fig. S6 is fitted well by a Gaussian function with FWHM (full width at half maximum) of 350 pm, which is essentially equivalent to that of pristine graphene (340 ± 15 pm). However, 30-min annealed graphene with the small dots shows an asymmetric distribution with much larger width. Decomposition with two Gaussians led to FWHM's of 390 and 600 pm. While the former is due to graphene itself, the latter can be attributed to the amorphous carbon nano-dots. These morphological changes are also consistent with the results obtained by the optical characterization represented by the variation of $I(D_{aC})$ in Fig. 2b.

**Discussion**

Our study shows that graphene is easily covered by hydrocarbons generated at elevated temperatures from polymers which are often unintentionally provided during the typical sample and device fabrication processes.[1,23] This demands extra caution in interpreting experimental results from graphene samples or devices that have undergone such processes. In particular, the presence of adhesive residues generated during micro-mechanical exfoliation has rarely been taken into account,[7,26] since the residues are typically not in direct contact with graphene flakes of interest. Through this study, however, it has been shown that such residues can adsorb on graphene flakes at elevated temperature. In this regard, it is also to be noted that isolated graphene flakes was contaminated by the polymeric residues dissolved and diffused by common solvents like acetone in a simple dipping process.[26] Some of the hydrocarbons may originate from the ambient air as Z. Li et al. recently showed in airborne contaminants-induced hydrophobicity of graphene.[27]

Furthermore, our study clearly shows that the polymeric adsorbates are not readily removed by brief thermal annealing, which is consistent with the recent observation by transmission electron microscopy (TEM).[24] As long as non-oxidizing, the nature of ambient gases in the heating chamber was found not to affect the efficiency of the removal greatly in contrast to the common preference for reducing atmosphere including hydrogen[9,28] rather than noble gases. In fact, the removal efficiency increased as decreasing the pressure inside the chamber from 1 atm to high vacuum, suggesting that gaseous species generated at elevated temperature adsorb on graphene. However, the observed pressure dependence is apparently very modest: a pressure decrease by several orders of magnitude led only to 50 ~ 70% decrease in $I(D_{aC})$. This suggests that graphene is also wetted by some fluidic adsorbates which migrate on the substrates, not necessarily airborne, at high temperature. Alternatively, this may be due to inefficient diffusion of the airborne species from the sample to the vacuum pump. Whereas even chemical species with high molecular mass can be evaporated from the sample located in the heating zone of the quartz tube, they will collide and condense on the less warm inner walls near the ends of the tube which remain near room temperature. The resulting large concentration gradient of the gaseous species along the tube axis will lead to reduced dependence of the efficiency on the base pressure which is measured at the cold end of the tube.

Finally, the current study provides a simple but reliable optical method to determine the amount of amorphous carbon. Although thermal degradation has been employed to remove various adsorbates on graphene samples, it has been not straightforward to confirm each cleanup process since the kinetics of decomposition varies widely from sample to sample according to the nature and amount of the foreign materials. Despite its preciseness,[24] for example, TEM cannot be applied to graphene supported on conventional thick substrates. As shown in the current report, quantification of the amount of aC with AFM is also subject to technical difficulties since the morphology of the aC overlayers can be indistinct depending on the specific details of thermal treatments, besides the fact that the method is time-consuming. In contrast, Raman spectroscopy can be carried out quickly without any extra preparation for samples. By simple comparison with a reference spectrum of pristine graphene, the amount of aC can be precisely determined through $I(D_{aC})$ and $I(G_{aC})$ at least on a relative scale. Using this analysis, it was demonstrated that aC originating from polymeric residues can be removed in both exfoliated and CVD-grown graphene by extended annealing in vacuum. This should be of practical importance to those who have to ensure surface cleanness of graphene.

In conclusion, we have investigated the Raman spectra of graphene annealed in inert gas or vacuum. In addition to the previously reported spectral shifts of G and 2D peaks, very broad additional Raman peaks were found near D, G and 2D frequencies. The new Raman peaks were attributed to amorphous carbon formed on the surface of graphene by carbonization of organic adsorbates originating from coexisting polymeric hydrocarbons in oxygen-free environment.

**Methods**

**Sample preparation.** One group of graphene samples were prepared by the micromechanical exfoliation method[1] using kish graphite (Covalent Materials Inc.) and Si substrates topped with 285 nm-thick $SiO_2$ layer. Another group of graphene samples were first grown on Cu foils by CVD and transferred onto the $Si/SiO_2$ substrates followed by chemical etching of the Cu foils.[23] Thin PMMA support layers were spin-coated on graphene before etching and removed by acetone after transfer.[23]

**Raman Spectroscopy.** The number of layers and structural quality of the samples were characterized by Raman spectroscopy.[18] Raman spectra were obtained with an Ar ion laser operated at 514.5 nm with average power of 1.5 mW focused onto a spot of ~0.5 μm in diameter by a 40 times objective with numerical aperture of 0.60. The spectral accuracy was found to be better than 1.0 cm$^{-1}$ with respect to Raman standards of graphite and Si crystals, whereas the full width at half maximum (FWHM) of the Rayleigh line was 6.0 cm$^{-1}$ using a spectrograph with focal length of 30 cm and a grating with 600 grooves/mm.[13] To obtain statistical significance, for each set of the experimental conditions, Raman spectra were obtained from 5 ~ 10 spots (1~2 μm apart) per each of a few graphene flakes larger than 10 μm along the long axis.

**Thermal treatments.** Thermal annealing consisted of linear ramping to a target temperature ($T_{anneal}$) in 30 min, holding temperature at $T_{anneal}$ for a given time ($\tau_{anneal}$), and cooling down to ~23 ºC for ~3 hours. To test the effects of gas environment in the quartz tube furnace, annealing was carried out in Ar gas (flow rate of 300 mL/min, ~1 atm), forming gas (5% $H_2$ in $N_2$, flow rate of 300 mL/min, ~1 atm), low vacuum (base pressure of ~$2 \times 10^{-3}$ Torr), and high vacuum (base pressure of ~3x10$^{-5}$ and ~1x10$^{-7}$ Torr). For oxidation, the temperature of samples was varied in the same manner as the thermal annealing, but in a mixed gas atmosphere (O$_2$:Ar = 350 mL/min: 700 mL/min, ~1 atm).

**Acknowledgements**

This work was supported by the National Research Foundation of Korea (No. 2012-053500, 2012-043136, 2012-0003059). The authors thank Hyeongkeun Kim and Sugang Bae for the access to high vacuum furnaces and Byung Hee Hong for providing CVD-grown graphene samples. The authors are also grateful to Chaiwon Kwon for providing carbon black powder and Taeg Yeoung Ko for helpful comments.

**Author contributions**

S.R. proposed and supervised the entire project. J.H., M.K.P., E.J.L., D.E.L and D.S.H. performed the experiments. J.H., M.K.P., E.J.L. and S.R. analysed the data. S.R. wrote the manuscript and all the authors participated in discussions of the research.

**Additional information**

Supplementary information accompanies this paper at http://www.nature.com/scientificreports

Competing financial interests: The authors declare no competing financial interests.

**Figure captions**

Figure 1. The effects of thermal annealing in single layer (1L) graphene. (a), (b) Optical micrograph of pristine and annealed graphene/SiO$_2$/Si, respectively ($T_{anneal}$ = 400 °C; $\tau_{anneal}$ = 30 min; in Ar gas). The yellow arrow points to polymeric adhesive residues. (c) Raman spectra of pristine (red) and annealed graphene with that of bare substrates (black). (d) Expanded view of (c).

Figure 2. The effects of various environmental parameters on the intensity of $D_{aC}$ peak, $I(D_{aC})$. (a) $I(D_{aC})$ of graphene obtained after annealing in various gas environment of the tube furnace ($T_{anneal}$ = 400 °C, $\tau_{anneal}$ = 2 hours). The first two measurements were carried out in Ar and forming gas of ~760 Torr, respectively, while the rest were in vacuum of the specified base pressure. $I(D_{aC})$ of each pristine sample is shown together for comparison. (b) $I(D_{aC})$ of graphene for varying $\tau_{anneal}$ ($T_{anneal}$ = 400 °C, in low vacuum). (c) $I(D_{aC})$ of exfoliated and CVD-grown graphene for varying $T_{anneal}$ ($\tau_{anneal}$ = 2 hours). The exfoliated graphene was annealed in low vacuum, while the CVD-grown graphene was in Ar gas. Each of data point in (a) ~ (c) represents a statistical average for 20 ~ 30 separate spectra from more than 3 graphene samples.

Figure 3. Topographic confirmation of aC layers on 2L graphene. (a) Optical micrograph of pristine 2L graphene. The dashed yellow square marks the position where the AFM images in (b) ~ (f) were obtained. The square was enlarged by a factor of two for visibility. (b) Non-contact AFM height image of the 2L graphene obtained after thermal oxidation at 550 °C for 4 hours, which generated 1L-deep (in the dotted circle) and 2L-deep (marked by the yellow arrow) etch pits. (c)~(f) AFM image obtained for varying $\tau_{anneal}$: (c) 1 , (d) 2, (e) 4, (f) 8 hours ($T_{anneal}$ = 700 °C; in Ar gas). The scanned area in (b) ~ (f) is 2x2 μm$^2$, respectively. Representative height profiles obtained along the white horizontal lines in (b) ~ (f) are given in Supplementary Fig. S5. (g) The height profile averaged over the yellow rectangle in (d). (h) $I(D_{aC})$ as a function of $\tau_{anneal}$. The orange dashed line represents the average $I(D_{aC})$ of pristine graphene.

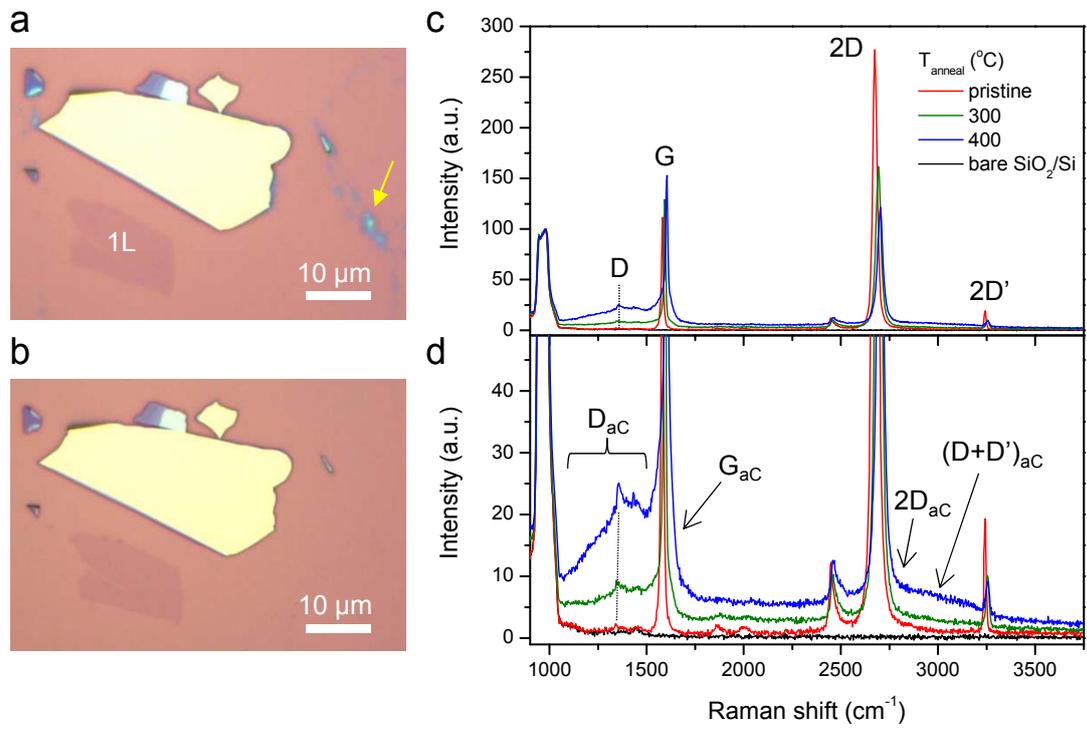

Fig.1. J. P. Hong et al.

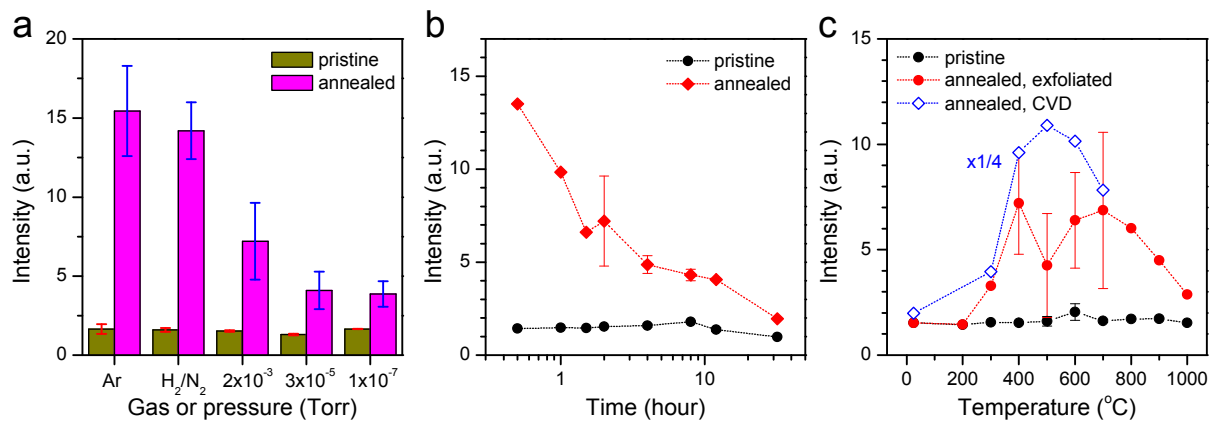

Fig.2. J. P. Hong et al.

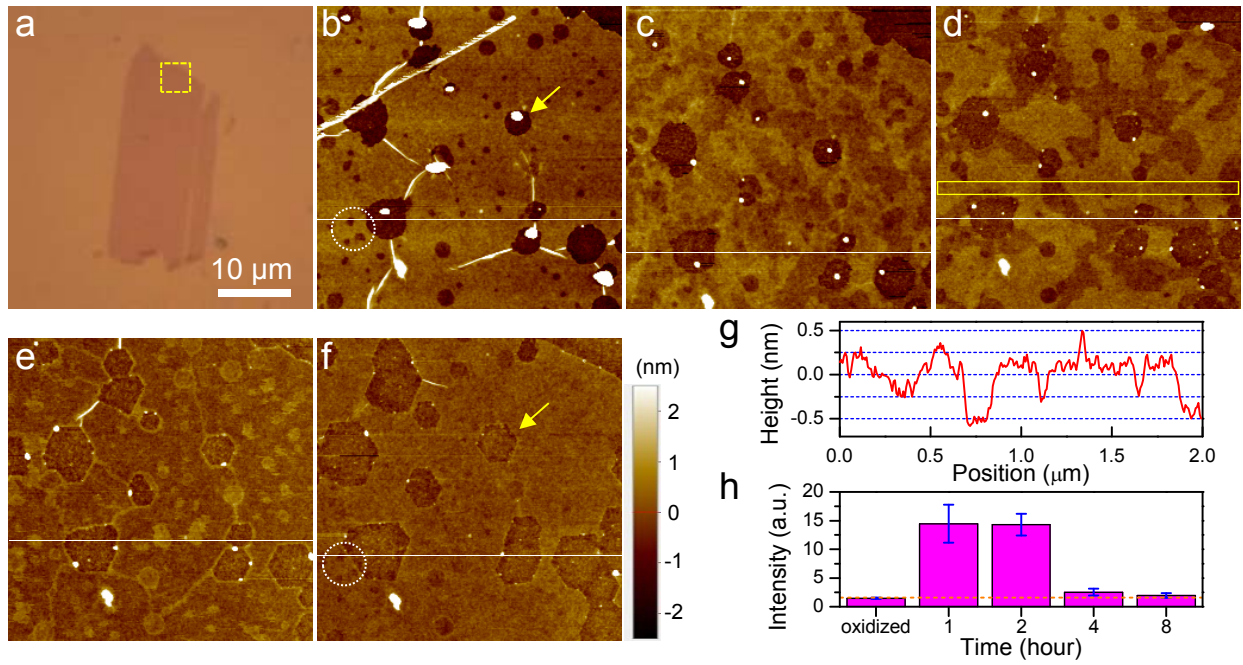

Fig.3. J. P. Hong et al.